\documentclass[english,12pt]{article}
\usepackage{array}
 \usepackage[normalem]{ulem}
\usepackage{graphicx}
\usepackage{amssymb}
\usepackage[english]{babel}
\usepackage{amsmath}
\usepackage{multirow}
\usepackage{prettyref}
\usepackage{babel}
\usepackage{units}
\usepackage[latin1]{inputenc}
\usepackage{amsfonts}
\usepackage{amssymb}
\usepackage[symbol]{footmisc}
\usepackage{babel}
\usepackage{color}
\usepackage[dvipsnames]{xcolor} 
\usepackage{cite}
\def\@fmsl@sh#1#2#3{\m@th\ooalign{$\hfil#1\mkern#2/\hfil$\crcr$#1#3$}}
 \def\eq#1\en{\begin{equation}#1\end{equation}}
\def\s[#1,#2]{[#1\stackrel{\star}{,}#2]}
\def\sx[#1,#2]{[#1\stackrel{\star_{x}}{,}#2]}

\newcommand{\GN}{{G_{\rm N}}}

\newcommand{\TP}{{t_{\rm P}}}

\def\gsim{\mathrel{\rlap{\lower4pt\hbox{\hskip1pt$\sim$}}
		\raise1pt\hbox{$>$}}}       

\newcommand{\gn}{G_{\rm N}}

\newcommand{\nc}{\newcommand}
\nc{\beq}{\begin{equation}}
\nc{\eeq}{\end{equation}}
\nc{\beqa}{\begin{eqnarray}}
\nc{\eeqa}{\end{eqnarray}}

\def\bc{\begin{center}}
\def\ec{\end{center}}

\def\to{\rightarrow}

\def\gsim{\mathrel{\mathpalette\atversim>}}

\def\bc{\begin{center}}
\def\ec{\end{center}}

\def\gsim{\mathrel{\rlap{\lower4pt\hbox{\hskip1pt$\sim$}}

    \raise1pt\hbox{$>$}}}       

\def\gsim{\mathrel{\rlap{\lower4pt\hbox{\hskip1pt$\sim$}}
    \raise1pt\hbox{$>$}}}       

\begin{document}
\makeatletter
\def\fmslash{\@ifnextchar[{\fmsl@sh}{\fmsl@sh[0mu]}}
\def\fmsl@sh[#1]#2{%
  \mathchoice
    {\@fmsl@sh\displaystyle{#1}{#2}}%
    {\@fmsl@sh\textstyle{#1}{#2}}%
    {\@fmsl@sh\scriptstyle{#1}{#2}}%
    {\@fmsl@sh\scriptscriptstyle{#1}{#2}}}
\def\@fmsl@sh#1#2#3{\m@th\ooalign{$\hfil#1\mkern#2/\hfil$\crcr$#1#3$}}
\makeatother

\thispagestyle{empty}
\begin{titlepage}
\boldmath
\begin{center}
  \Large {\bf   Quantum Gravitational Corrections in Cosmology}
    \end{center}
\unboldmath
\vspace{0.2cm}
\begin{center}
{\large Xavier Calmet,}$^{1,\,}$\footnote[1]{x.calmet@sussex.ac.uk}
{\large Roberto Casadio,}$^{2,\,3,\,4,\,}$\footnote[2]{casadio@bo.infn.it}
 {\large and}  
{\large Marco~Sebastianutti}$^{1,\,}$\footnote[3]{m.sebastianutti@sussex.ac.uk}
 \end{center}
\begin{center}
$^1${\sl Department of Physics and Astronomy, 
University of Sussex, Brighton, BN1 9QH, United Kingdom
}
\\
$^2${\sl Dipartimento di Fisica e Astronomia, Universit\`a di Bologna, via Irnerio 46, 40126 Bologna, Italy}
\\
$^3${\sl I.N.F.N., Sezione di Bologna, IS - FLAG, via B.~Pichat~6/2, 40127 Bologna, Italy}
\\
$^4${\sl Alma Mater Research Center on Applied Mathematics - AM$^2$, via Saragozza 8, 40123 Bologna, Italy}
\end{center}
\vspace{5cm}
\begin{abstract}
\noindent
We show how to reliably calculate quantum gravitational corrections to cosmological models using the unique effective action formalism for quantum gravity. Our calculations are model independent and apply to any ultra-violet complete theory of quantum gravity that admit general relativity as a low energy theory. We show that it is important to consider the full effective action to obtain renormalization group invariant solutions. We investigate the validity range of our techniques within simple cosmological models.
\end{abstract}  
\end{titlepage}



\newpage
\renewcommand{\thefootnote}{\arabic{footnote}}
\section{Introduction}
In this paper we use the unique effective action of quantum gravity \cite{Barvinsky:1983vpp,Barvinsky:1985an,Barvinsky:1987uw,Barvinsky:1990up,Buchbinder:1992rb,Calmet:2018elv} to calculate quantum gravitational corrections to several cosmology models.\footnote{The derivation of an effective action involves the splitting of the metric $g =\bar g+h$ in to a background $\bar g$ and a perturbation $h$. The perturbation $h$ is then integrated out. To perform this $h$-integration,
one must fix a gauge. In general, this gauge choice will also fix the gauge for $\bar g$ and it thus results in an effective action that is gauge dependent. This problem can be resolved using DeWitt's method of mean-field gauges. A further complication that arises is that the off-shell effective action in gauge theories depends on the choice of parameterization of the quantum fields. A solution to this issue was proposed in \cite{Vilkovisky:1984st}, leading to the unique effective action. For gravity, this effective action is constructed using covariant perturbation theory or the generalized Schwinger-DeWitt technique.} There have been previous studies of quantum gravitational effects in cosmology using the concept of effective action but these studies have limited themselves to either the local part of the effective action \cite{Codello:2015pga} or its non-local part \cite{Espriu:2005qn,Cabrer:2007xm,Donoghue:2014yha,Codello:2015pga} and these results are thus inconsistent as they are not renormalization group invariant. In \cite{Calmet:2023met}, the collapse of a dust ball has been considered using the Oppenheimer-Snyder model. The interior metric is given by the Friedmann-Lema\^itre-Robertson-Walker (FLRW) metric. This paper obtained for the first time quantum gravitational corrections to the FLRW metric which are renormalization group invariant.  Note that in this model the classical singularity is in the future and these results are thus not directly applicable to standard cosmological models.  The aim of our new paper is to extend these results to cosmology to demonstrate how to calculate reliably quantum gravitational corrections to cosmological models and to discuss their applicability domain.  

\section{A lightning review of the unique effective action}

The quantum corrections to classical solutions of general relativity are reliably calculable using quantum corrected field equations obtained from the variation of the unique effective action of quantum gravity. This effective action is obtained by integrating out the fluctuations of the graviton (and any other massless fields). It is formulated in terms of a curvature expansion.  At second order in curvature, the effective action is given by \cite{Barvinsky:1983vpp,Barvinsky:1985an,Barvinsky:1987uw,Barvinsky:1990up,Buchbinder:1992rb,Calmet:2018elv}
\begin{equation}
    \Gamma_{\rm QG} = \Gamma_{\rm L} + \Gamma_{\rm NL} +\Gamma_{\rm matter}
\end{equation}
with a local part
\begin{align}
    \Gamma_{\rm L} 
    &= 
    \int d^4x \, \sqrt{|g|} \left[ \frac{M_P^2 }{2}
    \big(\mathcal{R} - 2\Lambda\big)
    + c_1(\mu) \, \mathcal{R}^2 
    + c_2(\mu) \, \mathcal{R}_{\mu\nu} \mathcal{R}^{\mu\nu} 
    \right.\nonumber\\
    &\qquad \qquad \qquad \qquad
    + c_3(\mu) \, \mathcal{R}_{\mu\nu\rho\sigma} \mathcal{R}^{\mu\nu\rho\sigma} 
    + c_4(\mu) \, \Box \mathcal{R}
    + \mathcal{O}(M_P^{-2}) \Big]
\end{align}
and a non-local part 
\begin{align}
    \Gamma_{\rm NL} &= 
    - \int d^4x \, \sqrt{|g|} \left[ 
   \alpha \, \mathcal{R} \ln \left(\frac{\Box}{\mu^2} \right) \mathcal{R}
    +\beta \, \mathcal{R}_{\mu\nu} \ln \left( \frac{\Box}{\mu^2} \right) \mathcal{R}^{\mu\nu}
    \right.\nonumber\\
    &\qquad \qquad \qquad \qquad \left.
    +\gamma \, \mathcal{R}_{\mu\nu\rho\sigma} \ln \left( \frac{\Box}{\mu^2} \right) \mathcal{R}^{\mu\nu\rho\sigma}
    + \mathcal{O}(M_P^{-2}) \right],
\end{align}
and where $\Gamma_{\rm matter}$ describes the energy-momentum tensor of the cosmological model under consideration. We use the following standard notations: $M_P=\sqrt{\hbar \, c/(8 \,\pi \, G_N)}= 2.4\times10^{18}\,{\rm GeV}$ denotes the reduced Planck mass, $\Lambda$ the cosmological constant (which we set to zero for simplicity) and $c_i$ and $\alpha$, $\beta$ and $\gamma$ are Wilson coefficients. 

While the Wilson coefficients of the non-local part of the action are calculable from first principles and given in Table 1, those of the local part of the action remain undetermined from the unique effective action point of view.  Furthermore, the parameter $\mu$ is a renormalization scale.
\begin{table}
\center
\begin{tabular}{| c | c | c | c |}
\hline
 & $ \alpha $ & $\beta$ & $\gamma$  \\
 \hline
 \text{Scalar} & $ 5(6\xi-1)^2$ & $-2 $ & $2$     \\
 \hline
 \text{Fermion} & $-5$ & $8$ & $7 $ \\
 \hline
 \text{Vector} & $-50$ & $176$ & $-26$ \\
 \hline
 \text{Graviton} & $250$ & $-244$ & $424$\\
 \hline
\end{tabular}
\caption{Non-local Wilson coefficients for different fields.
All numbers should be divided by $11520\pi^2$. Here, $\xi$ denotes the value of the non-minimal coupling for a scalar theory.}
\label{coeff1}
\end{table}

The renormalization scale $\mu$ in the effective action is a free parameter. We note that physical observables should not depend on the renormalization scale. The renormalization group equations for the Wilson coefficients of the local part of the effective action are given by
\begin{align}
c_1(\mu) &= c_1(\mu_\ast) - \alpha \, \ln\left( \frac{\mu^2}{\mu_\ast^2}\right), \\
c_2(\mu) &= c_2(\mu_\ast) - \beta \, \ln\left( \frac{\mu^2}{\mu_\ast^2}\right), \\
c_3(\mu) &= c_3(\mu_\ast) - \gamma \, \ln\left( \frac{\mu^2}{\mu_\ast^2}\right),
\end{align}
where $\mu_\ast$ is the scale at which the effective action is matched to the ultra-violet complete theory of quantum gravity \cite{Buchbinder:1992rb}.

It was shown in  \cite{Han:2004wt,Atkins:2010eq,Atkins:2012yn,Calmet:2014gya} that in models with a large number of fields, the scale at which quantum gravitational effects become strong depends on the number of fields in the model and the non-minimal coupling to curvature of scalar fields.  The effective Planck mass is given by $M_P\sqrt{160\pi/N}$ where $N=1/3 N_S+ N_F+4 N_V$ ($N_S$, $N_F$ and $N_V$ are the numbers of real scalar fields, fermions and vector fields in the model). A large non-minimal coupling of scalar fields to curvature reduce the effective Planck mass to $M_P/\xi$. Furthermore,  the local Wilson coefficients are expected to be of order $N$ because of their renormalization group equations.

The quantum gravitational field equations to second order in curvature can be derived from the unique effective action. They are given by  
	\begin{eqnarray} \label{FEq}
		{\cal R}_{\mu\nu} - \frac{1}{2}\, {\cal R}\, g_{\mu\nu} - 16\,\pi\,\gn  \left( H_{\mu\nu}^{\rm L} + H_{\mu\nu}^{\rm NL} \right)
		= 8 \, \pi \, G_{\rm N} T_{\mu\nu}
		\ ,
	\end{eqnarray}
		where $G_{\rm N}$ is Newton's constant, $T_{\mu\nu}$ is the energy-momentum tensor,
	\begin{align}
		H_{\mu\nu}^{\rm L} 
		=
		&\,  
		\bar{c}_1
		\left( 2\, {\cal R}\, {\cal R}_{\mu\nu} - \frac{1}{2}\, g_{\mu\nu}\, {\cal R}^2 + 2\, g_{\mu\nu}\, \Box {\cal R} - 2 \nabla_\mu \nabla_\nu {\cal R}\right) 
		\label{eq:EQMLoc}
		\\
		&\,
		+\bar{c}_2
		\left( 2\, {\cal R}_{~\mu}^\alpha\, {\cal R}_{\nu\alpha} - \frac{1}{2}\, g_{\mu\nu}\, {\cal R}_{\alpha\beta}\, {\cal R}^{\alpha\beta}
		+ \Box {\cal R}_{\mu\nu} + \frac{1}{2}\, g_{\mu\nu}\, \Box {\cal R}
		- \nabla_\alpha \nabla_\mu {\cal R}_{~\nu}^\alpha
		- \nabla_\alpha \nabla_\nu {\cal R}_{~\mu}^\alpha \right) ,
		\nonumber
	\end{align}
	with $\bar{c}_1=c_1-c_3$, $\bar{c}_2=c_2+4c_3$ and
	\begin{align}
		H_{\mu\nu}^{\rm NL} 
		=
		&\,
		- 2\,\alpha
		\left( {\cal R}_{\mu\nu} - \frac{1}{4}\, g_{\mu\nu}\, {\cal R}
		+ g_{\mu\nu}\, \Box
		- \nabla_\mu \nabla_\nu \right)
		\ln\left(\frac{\Box}{\mu^2}\right)\, {\cal R}
		\nonumber
		\\
		&\, 
		- \beta
		\bigg( 2\, \delta_{(\mu}^\alpha\, {\cal R}_{\nu)\beta}
		- \frac{1}{2}\, g_{\mu\nu}\, {\cal R}_{~\beta}^\alpha
		+ \delta_{\mu}^\alpha\, g_{\nu\beta}\, \Box
		+ g_{\mu\nu}\, \nabla^\alpha \nabla_\beta  \nonumber  \\ \nonumber 
		&\quad 
		- \delta_\mu^\alpha\, \nabla_\beta \nabla_\nu
		- \delta_\nu^\alpha\, \nabla_\beta \nabla_\mu 
		\bigg)
		\ln\left(\frac{\Box}{\mu^2}\right)\, {\cal R}_{~\alpha}^\beta &
		\nonumber
		\\
		&\,
		- 2 \,\gamma
		\left[
		\delta_{(\mu}^\alpha\, {\cal R}_{\nu)~\sigma\tau}^{~\beta}
		- \frac{1}{4}\, g_{\mu\nu}\, {\cal R}^{\alpha\beta}_{~~\sigma\tau}
		+\left( \delta_\mu^\alpha\, g_{\nu\sigma} + \delta_\nu^\alpha\, g_{\mu\sigma} \right)
		\nabla^\beta \nabla_\tau \right]
		\ln\left(\frac{\Box}{\mu^2}\right)\, {\cal R}_{\alpha\beta}^{~~\sigma\tau}
		\ .
	\end{align}
	
If the physical system under consideration has weak space-time curvature, perturbation theory can be applied to solve these coupled partial differential equations. We can obtain a controlled approximation by using  perturbation theory around the classical solution. We  set $g_{\mu\nu} = g_{\mu\nu}^{\rm c} + g_{\mu\nu}^{\rm q}$ where $g_{\mu\nu}^{\rm c}  $ is the classical solution and  $g_{\mu\nu}^{\rm q}$ the quantum solution one is solving Eq. \eqref{FEq} for. 

\section{Quantum gravitational corrections to FLRW cosmology}
In the spatially flat FLRW cosmological models, the metric is defined by the line element
	\begin{eqnarray}
    ds^2=dt^2-a(t)^2\left(dx^2+dy^2+dz^2\right),
\end{eqnarray}
$t$ being the cosmic proper time, $a(t)$ the scale factor and $\{x,y,z\}$ the spatial comoving coordinates. The matter content filling this universe is taken to be a perfect fluid with an equation of state $p=w\rho$.
For such a classical background, the 00-component of the Einstein equations is the Friedmann constraint,
\begin{eqnarray}\label{eq:FriedmannC}
    H_c^2=\frac{8\pi G_{\rm N}}{3}\rho,
\end{eqnarray}
(where the subscript stands for classical) which must hold along with the continuity equation
\begin{eqnarray}
    \dot{\rho}=-3H(1+w)\rho,
\end{eqnarray}
which is obtained from imposing the conservation of energy and momentum, i.e. $\nabla_{\mu}T^{\mu\nu}=0$. The classical solutions for the scale factor $a(t)$ and the Hubble function $H(t)\equiv\dot{a}/a$ read:
\begin{align}
    a_c(t)&=a_0\left(\frac{t}{t_0}\right)^{\frac{2}{3(1+w)}},\label{eq:aC}
    \\
    H_c(t)&=\frac{2}{3(1+w)\,t},
\end{align}
where $a_0$ and $t_0$ are arbitrary initial conditions and $t\in[0,+\infty]$ for expanding solutions (we will also consider contracting ones in the following). We will focus mainly on non-relativistic matter or dust for which $w=0$ and $a(t)\sim t^{2/3}$ but we will also briefly discuss the case of radiation for which $w=1/3$ and $a(t)\sim t^{1/2}$ and on a cosmological constant component $w=-1$ characterised by $H=\sqrt{\Lambda/3}=\text{const.}$ and a scale factor
\begin{eqnarray}
    a_c(t)=a_0 e^{H\left(t-t_0\right)}.
\end{eqnarray}

For FLRW spacetimes the Weyl tensor $C_{\mu\nu\rho\sigma}=0$, hence $H_{\mu\nu}^{\rm L}$ and $H_{\mu\nu}^{\rm NL}$ simplify to 
\begin{eqnarray}
    H_{\mu\nu}^{\rm L}&=&\hat{c}_1(\mu)\left(-\frac{1}{2}g_{\mu\nu}R^2+2R_{\mu\nu}R-2\nabla_{\mu}\nabla_{\nu}R+2g_{\mu\nu}\Box R\right),\label{eq:HL}\\
    H_{\mu\nu}^{\rm NL}&=&-\hat{\alpha}\left(-\frac{1}{2}g_{\mu\nu}R+2R_{\mu\nu}-2\nabla_{\mu}\nabla_{\nu}+2g_{\mu\nu}\Box \right)\log{\left(\frac{\Box}{\mu^2}\right)}R.
    \label{eq:HNL}
\end{eqnarray}
where $\hat{c}_1=c_1+\frac{1}{3}(c_2+c_3)$, and $\hat{\alpha}=\alpha+\frac{1}{3}(\beta+\gamma)$.

In order to find the explicit form of the quantum corrections to the scale factor $a(t)$ (and the Hubble function $H(t)$), we take the 00-component of~\eqref{FEq}, which is a quantum corrected version of the Friedmann constraint equation~\eqref{eq:FriedmannC}, and solve it using perturbation theory around the classical solutions i.e. $a(t)\rightarrow a_c(t)+ a_q(t)$. We find
\begin{eqnarray}\label{eq:EOMsPlot}
    H^2+\frac{16\pi\GN}{3}\left[H_{00}^{\rm L}(a_c)+H_{00}^{\rm NL}(a_c)\right]=\frac{8\pi\GN}{3}\rho,
\end{eqnarray}
where $H^2=H_c^2+2H_c H_q+{\cal O}(H_q^2)$ with 
\begin{eqnarray}\label{eq:deltaHdeltaa}
    H_q=a_c^{-1}\left(\dot a_q-H_c a_q \right)+{\cal O}(a_q^2),
\end{eqnarray}
and substituting $\GN=\TP^2$, $\TP$ being the Planck time, we arrive at
\begin{eqnarray}\label{eq:deltaH}
    H_q=-\frac{16\pi\TP^2}{6H_c}\left(H_{00}^{\rm L}+H_{00}^{\rm NL}\right).
\end{eqnarray}

The local term in~\eqref{eq:HL}, evaluated on a spatially-flat FLRW background with $w\neq-1$, is given by
\begin{eqnarray}
    H_{00}^{\rm L}&=&18\hat{c}_{1}(\mu)\left(6H_{c}^2\dot{H}_{c}-\dot{H}_{c}^2+2H_{c}\ddot{H}_{c}\right),
    \nonumber
    \\
    &=&-\frac{8(1-3w)}{(1+w)^3}\frac{\hat{c}_{1}(\mu)}{t^4}.
\end{eqnarray}
While it vanishes for $w=1/3$ and  for $w=-1$,  for $w=0$ we have
\begin{eqnarray}
    H_{00}^{\rm L}=-\frac{8\hat{c}_{1}(\mu)}{t^4}.
\end{eqnarray}

We now compute the non-local term $H_{00}^{\rm NL}$. In the case $w=0$, we distinguish between an expanding and a collapsing flat FLRW universe with time domain $t\in[0,+\infty]$ and $t\in[-\infty,0]$ for the expanding and collapsing case respectively. Note that $H_{00}^{\rm L}=0$ and $H_{00}^{\rm NL}=0$ for $w=1/3$ and $w=-1$ only holds to second order in the curvature expansion. They do not vanish at higher order in this expansion.

Using the representation given in \cite{Frob:2012ui} for the distribution $\log \Box$, one has 
\begin{eqnarray}
    F(t)&=&\log{\left(\frac{\Box}{\mu^2}\right)}R(x)=\int d^4y\left[\frac{\sqrt{-g(y)}}{\sqrt{-g(x)}}\right]^{\frac{1}{2}}L(x-y)R(y),
    \nonumber
    \\
    &=&a(t)^{-\frac{3}{2}}\int dt^\prime a(t^\prime)^{\frac{3}{2}}L(t-t^\prime)R(t^\prime),
\end{eqnarray}
where 
\begin{eqnarray}
L(t-t^\prime)= -2 \lim_{\epsilon \to 0}  \left [ \log \left(\mu_{\rm E} \epsilon \right) \delta(t-t^\prime)+\frac{\Theta\left(t-t^\prime-\epsilon\right)}{t-t^\prime} \right ],
\end{eqnarray}
and 
$\mu_{\rm E}=\mu\cdot e^{\gamma_{\rm E}}$, $\gamma_{\rm E}$ being the Euler constant.

For radiation, $w=1/3$, and for a cosmological constant, $w=-1$, the non-local quantum corrections vanish. Note that this was pointed out in \cite{Codello:2015pga} previously. Note as well that our results differ from those obtained in \cite{Espriu:2005qn} and \cite{Cabrer:2007xm} as we used a different expansion of the non-local function $L(t-t^\prime)$ which is consistent with our unique effective action approach.

 For an expanding dust-filled flat FLRW universe, we find
\begin{eqnarray}
F(t)=\frac{8}{3t^2}\left[\log{\left(\frac{t}{t_i}-1\right)}+\log{\left(\mu_{\rm E}t\right)}\right], 
\end{eqnarray}
where $t_i$ corresponds to the lower integration cut-off to avoid integrating out over the singularity at $t=0$.

We then obtain
\begin{eqnarray}
    H_{00}^{\rm NL}&=&-\frac{16\hat{\alpha}}{t^4}\left[\frac{2}{3}\left(\frac{t}{t_i-t}-1\right)+\log{\left(\mu_{\rm E} t\right)}+\log{\left(\frac{t}{t_i}-1\right)}\right],
\end{eqnarray}
and substituting the above expressions for $H_{00}^{\rm L}$ and $H_{00}^{\rm NL}$ evaluated on the classical solution in~\eqref{eq:deltaH} we find:
\begin{eqnarray}
    H(t)&=&\frac{2}{3t}+\frac{32\pi\TP^2\hat{c}_1(\mu)}{t^3}+\frac{64\pi\TP^2\hat{\alpha}}{t^3}\left[\frac{2}{3}\left(\frac{t}{t_i-t}-1\right)+\log{\left(\mu_{\rm E}t\right)}+\log{\left(\frac{t}{t_i}-1\right)}\right]
    \nonumber
    \\
&&    +{\cal O}( H_q^2).
 \label{eq:solH}
\end{eqnarray}
We find the following ODE for the leading order quantum correction to the  scale factor
\begin{eqnarray}
\label{eq:ODEa}
    \dot{a}_q-\frac{2}{3t}a_q
    =
    a_0\left(\frac{t}{t_0}\right)^{\frac{2}{3}}
    \frac{32\pi\TP^2}{t^3}
    \left\{
    \hat{c}_1(\mu)+2\,\hat{\alpha}\left[\frac{2}{3}\left(\frac{t}{t_i-t}-1\right)+\log{\left(\mu_{\rm E}t\right)}+\log{\left(\frac{t}{t_i}-1\right)}\right]\right\}. 
    \nonumber \\ &&
\end{eqnarray}
Its solution is given by
\begin{eqnarray}
    a(t)&=&a_0\left(\frac{t}{t_0}\right)^{\frac{2}{3}}\bigg\{1-\frac{16\pi\TP^2\hat{c}_1(\mu)}{t^2}+   \\ \nonumber  &&   +  
        \frac{16\pi\TP^2\hat\alpha}{t^2}\left[\frac{1}{3}\left(1-2\frac{t}{t_i}\right)-2\log{\left(\mu_{\rm E} t\right)}-2\log{\left(\frac{t}{t_i}-1\right)}-\frac{2}{3}\frac{t^2}{t_i^2}\log{\left(1-\frac{t_i}{t}\right)}\right]\bigg\} \\ \nonumber  &&   +   C_1t^{\frac{2}{3}}+{\cal O}(a_q^2),\label{eq:sola}
\end{eqnarray}
where $C_1$ is an arbitrary integration constant which we set to zero.

We emphasize that $a(t)$ is renormalization group invariant. The scale dependence of the non-local part is canceled by the scale dependence of the Wilson coefficient $\hat{c}_1(\mu)$. This can be checked explicitly using 
\begin{eqnarray}\label{eq:running}
    \hat{c}_1(\mu)=\hat{c}_1(\mu_*)-2\hat{\alpha}\log{\left(\frac{\mu}{\mu_*}\right)}.
\end{eqnarray}
Note that this is the main result of this work, previous works had considered the local or non-local part, but this is the first one to emphasize that a consistent quantum gravitational correction can only be obtained by adding the local and non-local corrections. The solution obtained using our framework is renormalization group invariant and thus physically relevant. 

The value of the non-local coefficient $\hat{\alpha}$ depends on the number of fields $N_i$ contributing to the graviton self energy where the index $i$ can stand for scalar ($S$), fermion ($F$), vector ($V$) and spin-2 ($G$) fields:
\begin{eqnarray}
    \hat{\alpha}&=&\frac{1}{11520\pi^2}\left(N_S\hat{\alpha}_{S}+N_{F}\hat{\alpha}_{F}+N_V\hat{\alpha}_V+N_G\hat{\alpha}_G\right),\nonumber\\
    &=&\frac{1}{11520\pi^2}\left(\,5(6\xi-1)^2N_S+310N_G\right),\label{eq:alpha2}
\end{eqnarray}
where $\xi$ is the non-minimal coupling between the scalar field and the Ricci scalar, the values of $\hat{\alpha}_i$  are calculated using the values of $\alpha_i$, $\beta_i$ and $\gamma_i$ given in Table 1. In particular, we note that $\hat{\alpha}_{F}=\hat{\alpha}_V=0$, hence fermions and photons do not affect the value of $\hat{\alpha}$ which, independently of the scalar field coupling $\xi$ and their number $N_S$, is always positive. In general relativity, $N_G=1$.

For a collapsing dust-filled flat FLRW universe, we obtain
\begin{eqnarray}
    H(t)=\frac{2}{3t}+\frac{32\pi\TP^2\hat{c}_1(\mu_*)}{t^3}+\frac{64\pi\TP^2\hat{\alpha}}{t^3}\left[-\frac{2}{3}+\log{\left(-\mu_{\rm E}^{*}t\right)}\right]+{\cal O}(H_q^2).
\end{eqnarray}
and
\begin{eqnarray}
    a(t)=a_0\left(\frac{t}{t_0}\right)^{\frac{2}{3}}\bigg\{1-\frac{16\pi\TP^2\hat{c}_1(\mu_*)}{t^2}+\frac{16\pi\TP^2\hat\alpha}{t^2}\left[\frac{1}{3}-2\log{\left(-\mu_{\rm E}^{*} t\right)}\right]\bigg\}+C_1t^{\frac{2}{3}}+{\cal O}(a_q^2),
\end{eqnarray}
where $C_1$ is an integration constant which we again set to zero.

Note that  we have performed two expansions: a curvature expansion at the level of the effective action  and a perturbative expansion around the classical solution, i.e. an expansion in $\hbar$. We are thus limited by two criteria in terms of validity of our approximations, we need to impose that $t> 10/(\sqrt{160\pi/N}) t_P$ (it has been shown in \cite{Calmet:2021lny} that the semi-classical regime can be trusted up to $\sim M_P(\sqrt{160\pi/N})/10$). Furthermore, we need to require that $\left|a_q(t)\right|<\left|a_c(t)\right|$ (equivalently $\left|H_q(t)\right|<\left|H_c(t)\right|$) because of the perturbative theory used to solve the ODE.

\begin{figure}[ht!]
\centering 
\includegraphics[width=0.32\textwidth]{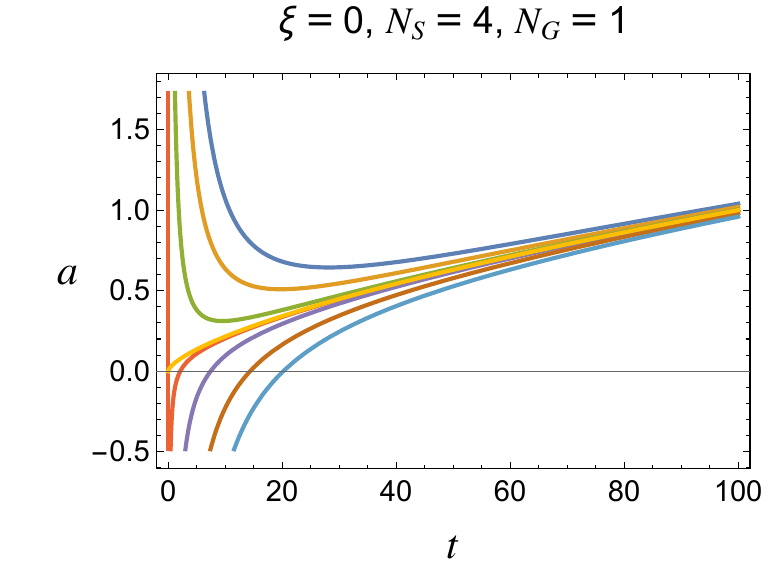}
\includegraphics[width=0.32\textwidth]{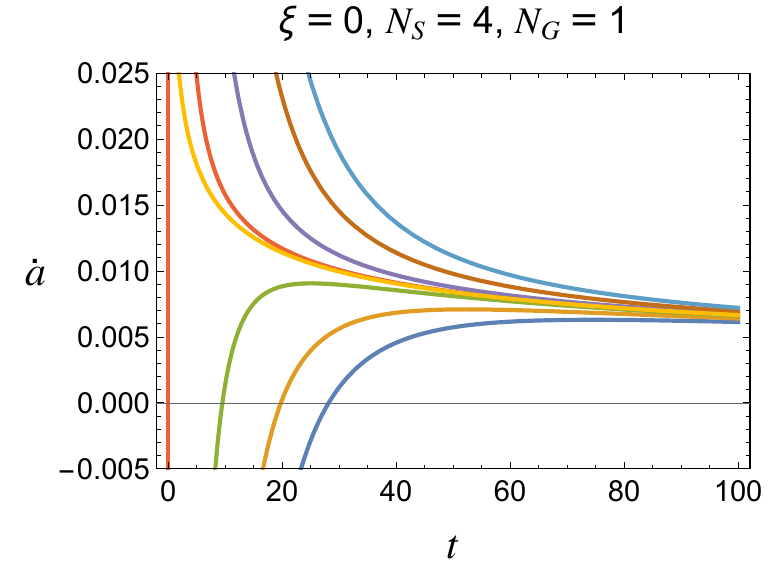}
\includegraphics[width=0.32\textwidth]{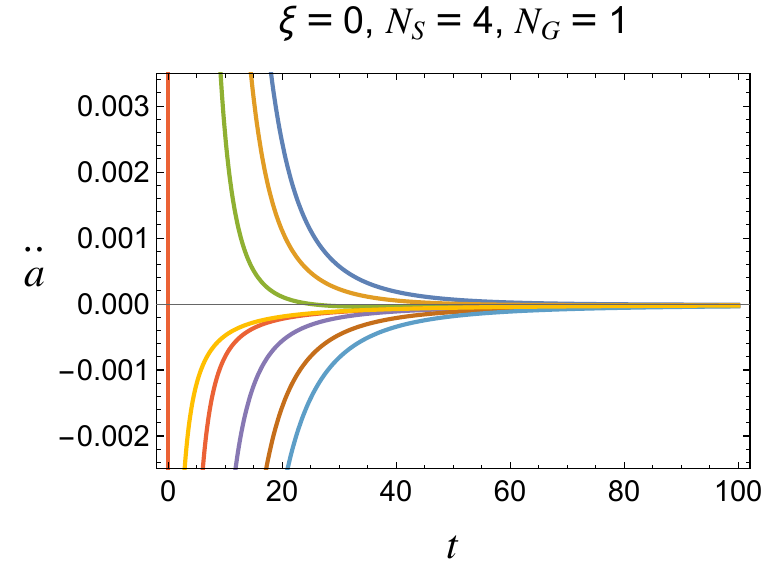}
\includegraphics[width=0.52\textwidth]{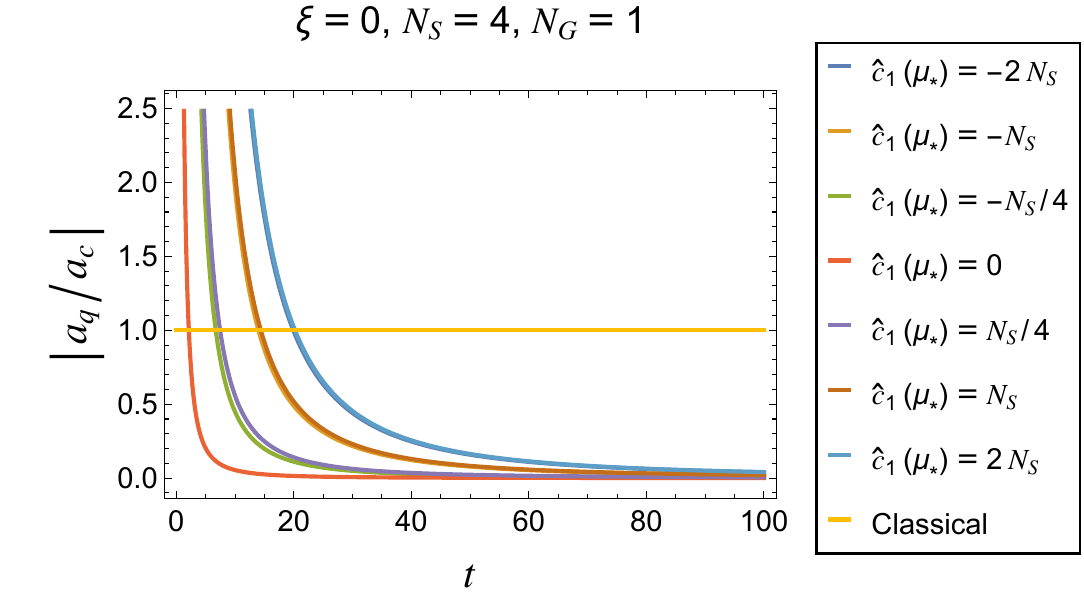}
\caption{Expanding spatially-flat dust-filled FLRW universe.}
\label{fig:expansion}
\end{figure}

\begin{figure}[ht!]
\centering 
\includegraphics[width=0.32\textwidth]{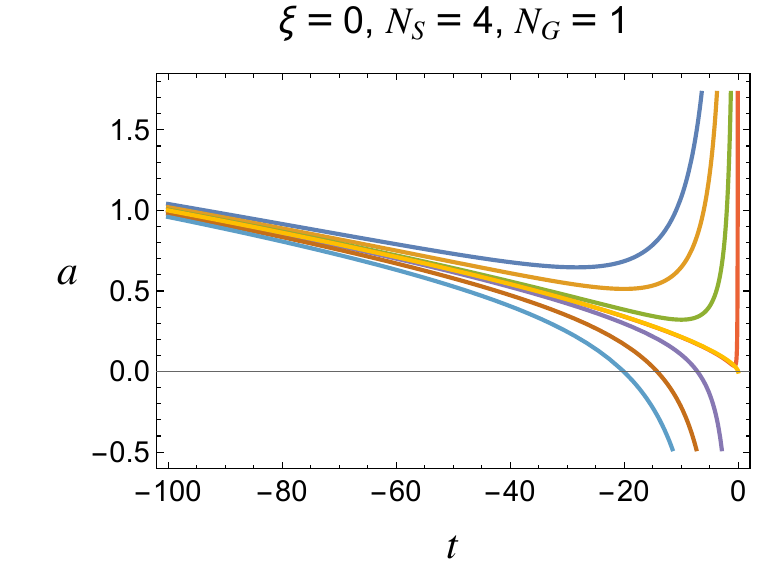}
\includegraphics[width=0.32\textwidth]{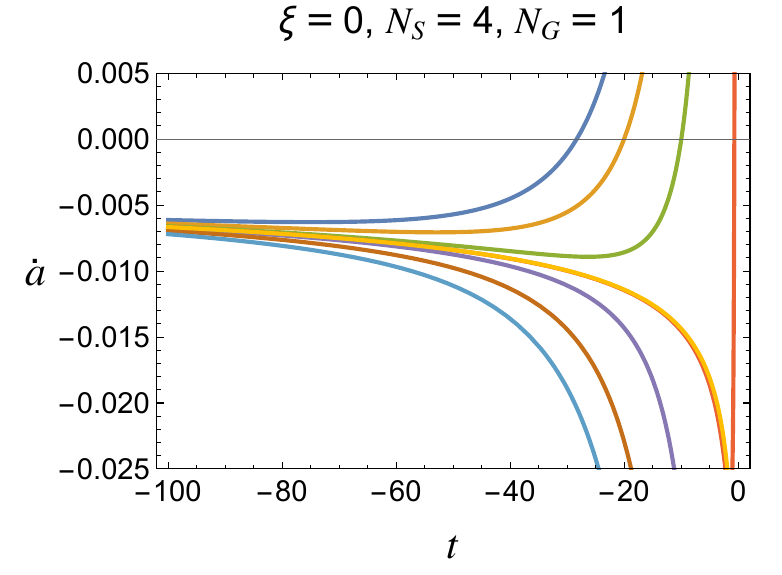}
\includegraphics[width=0.32\textwidth]{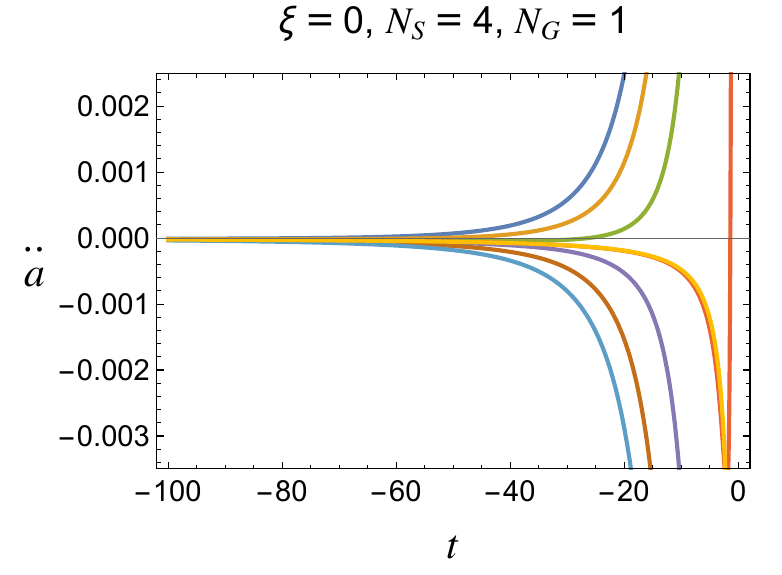}
\includegraphics[width=0.52\textwidth]{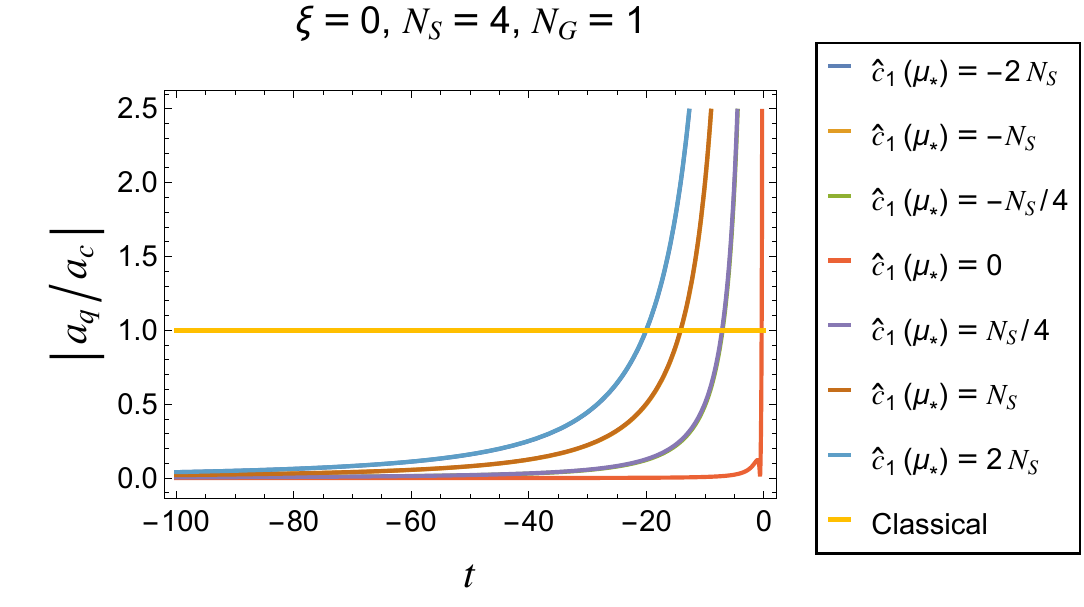}
\caption{Collapsing spatially-flat dust-filled FLRW universe.}
\label{fig:collapse}
\end{figure}

We now plot the scale factors for an expanding, Fig.~\ref{fig:expansion},  and collapsing, Fig.~\ref{fig:collapse}, dust-filled flat FLRW universe.
We vary $\hat{c}_1(\mu_*)$ and fix the parameters: $\mu_{\rm E}^*=\mu_* \cdot e^{\gamma_{\rm E}}\sim1.78\mu_*$, $C_1=\xi=0$, $N_S=4$, $N_G=1$, which are the Standard Model values, $(\mu_*)^{-1}=M_P\sqrt{160\pi/N}$, $a_0=1$ and $\TP=\pm1$, $t_0=\pm100$, for the expanding and contracting case respectively. Note that times are in Planck time units.

The difference between the expanding and collapsing universes arises from the non-local integral. While for the collapsing FLRW universe we are not integrating over the singularity at $t=0$ and thus do not need to cut-off the integral, in the expanding FLRW we need to introduce a cut-off $t_i$ to avoid integrating over the singularity at $t=0$. In these plots we have set $t_i=10^{-6}$ to cut-off the singularity.

In both figures, we display beside the scale factor $a(t)$, its first and second time derivatives $\dot{a}(t)$ and $\ddot{a}(t)$: while the former enables us to detect a bounce that can happen at $t=t_b$ for which $\dot{a}(t_b)=0$, the latter enables us to check if a phase of accelerated expansion can originate just from the quantum corrections to a spatially-flat dust-filled FLRW universe.

In an expanding universe, for negative values of $\hat{c}_1(\mu_*)$, we observe both a bounce and an early-time phase of positive acceleration of the universe. While the former lays outside the regime of validity of perturbation theory as can be see from the $|a_q/a_c|$ plot, the latter phase seems to happen when the ratio $|a_q/a_c| \ll1$, hence when perturbation theory can still be trusted. At later times the quantum corrections become small and the acceleration converges to the negative value given by the classical dust-filled case. A collapsing universe has an analogous behaviour for negative values of $\hat{c}_1(\mu_*)$: we observe both a bounce and a late-time phase of positive acceleration. However, this happens in a time region where we cannot trust our approximations.

The validity range of our approximations is constrained by the requirement to remain in the semi-classical regime i.e.  $t> 10/(\sqrt{160\pi/N}) t_P\sim t_P$ for $N=4$. The validity of perturbation theory can be visualised in the plot of $\left|a_q/a_c\right|$ in
Fig.~\ref{fig:expansion} and Fig.~\ref{fig:collapse}. Whenever the quantum correction becomes of the same order as the classical solution, we need to stop trusting our approximative solution. These plots show that the validity range depends on the physical parameters of the universe under consideration.

\section{Conclusions}

In this paper we have shown how to reliably calculate quantum gravitational corrections to cosmological models. We have shown that it is important to consider the full effective action to obtain renormalization group invariant solutions. We have investigated the validity range of our techniques within simple cosmological models. Any large effect such as bounces cannot be described reliably within our approximations as by definition in such cases, the quantum corrections dominate over the classical solution. However, our results can be considered as tantalising hints that quantum corrections could indeed lead to bounces in the late-time cosmological models and accelerated expansion in the past of the universe. {Note that our work could easily be extended to a  multi-component  cosmological model or even models with a general perfect fluid with a variable equation of state. However, given the smallness of the effect, there is little value in doing actual phenomenology. The main contribution of our work is to show that quantum gravitational corrections in cosmology are calculable from first principles.}

{\it Acknowledgments:}
	The work of X.C.~is supported in part  by the Science and Technology Facilities Council (grants numbers ST/T006048/1 and ST/Y004418/1.).
	The work of R.C.~is partially supported by the INFN grant FLAG and my work has also been carried out in
the framework of activities of the National Group of Mathematical Physics (GNFM, INdAM).
	The work of M.S.~is supported by a doctoral studentship of the Science and Technology Facilities Council (training grant No. ST/X508822/1, project ref. 2753640).
\\
	\bigskip 
	
	{\it Data Availability Statement:}
	This manuscript has no associated data. Data sharing not applicable to this article as no datasets were generated or analysed during the current study.
	
	\bigskip 


\bigskip{}

\end{document}